\documentclass[prd,aps,floats,preprintnumbers,preprint]{revtex4}
\usepackage[dvipdfm]{graphicx,color}

\newcommand{\be}{\begin{equation}}
\newcommand{\ee}{\end{equation}}
\newcommand{\bea}{\begin{eqnarray}}
\newcommand{\eea}{\end{eqnarray}}

\begin{document}


\title{Effects of inhomogeneities on apparent cosmological observables:
``fake'' evolving dark energy}

\author{Antonio Enea Romano$^{1,2,3,5}$, Alexei A. Starobinsky$^{4,5}$, and Misao Sasaki$^1$}
\affiliation
{$^1$Yukawa Institute for Theoretical Physics, Kyoto University,
Kyoto 606-8502, Japan\\
$^2$Instituto de Fisica, Universidad de Antioquia, A.A.1226, Medellin, Colombia \\
$^3$Leung Center for Cosmology and Particle Astrophysics, National Taiwan University, Taipei 10617, Taiwan, R.O.C.\\
$^4$L. D. Landau Institute for Theoretical Physics,
Moscow 119334, Russia\\
$^5$Research Centre for the Early Universe (RESCEU), Graduate School of Science,
The University of Tokyo, Tokyo 113-0033, Japan\\
}

\begin{abstract}
Using the exact Lemaitre-Bondi-Tolman solution with a
non-vanishing cosmological constant $\Lambda$, we investigate how
the presence of a local spherically-symmetric inhomogeneity can
affect apparent cosmological observables, such as the deceleration
parameter or the effective equation of state of dark energy (DE),
derived from the luminosity distance under the assumption that the
real space-time is exactly homogeneous and isotropic. The presence
of a local underdensity is found to produce apparent phantom
behavior of DE, while a locally overdense region leads to apparent
quintessence behavior. We consider relatively small large scale inhomogeneities which today
are not linear and could be seeded by primordial curvature perturbations
compatible with CMB bounds. Our study shows how observations in an
inhomogeneous $\Lambda$CDM universe with initial conditions
compatible with the inflationary beginning, if interpreted under
the wrong assumption of homogeneity, can lead to the wrong
conclusion about the presence of ``fake'' evolving dark energy
instead of $\Lambda$.

\end{abstract}

\maketitle
\section{Introduction}

High redshift luminosity distance measurements
\cite{Perlmutter:1999np,Riess:1998cb,Tonry:2003zg,Knop:2003iy,Barris:2003dq,Riess:2004nr}
and the WMAP measurement \cite{WMAP2003,Spergel:2006hy} of cosmic
microwave background (CMB) interpreted in the context of standard
FLRW cosmological models have strongly disfavored a matter
dominated universe,
 and strongly supported a dominant dark energy component, giving rise
to a positive cosmological acceleration.
As an alternative to dark energy, it has been
proposed \cite{Nambu:2005zn,Kai:2006ws}
 that we may be at the center of an inhomogeneous isotropic universe
described by a Lemaitre-Tolman-Bondi (LTB) solution of Einstein's field
equations, where spatial averaging over one expanding and one contracting
region is producing a positive averaged acceleration $a_D$, but is has been shown
 how in general this procedure can lead to formal definition of unobservable
quantities \cite{Romano:2006yc}.
Another more general approach to map luminosity distance as a function of
 redshift $D_L(z)$ to LTB models has been
 proposed \cite{Chung:2006xh},
 showing that an inversion method can be applied successfully to
reproduce the observed $D_L(z)$. Interesting analysis of
observational data in inhomogeneous models without dark energy is
given for example in \cite{Alexander:2007xx,Alnes:2005rw}.

The main point is that the luminosity distance is in general sensitive
to the geometry of the space through which photons are propagating along
null geodesics, and therefore arranging appropriately the geometry of
a given cosmological model it is possible to reproduce a given $D_L(z)$.
For FLRW models this corresponds to the determination of $\Omega_{\Lambda}$
 and $\Omega_m$ and for LTB models it allows to determine the
functions $E(r),M(r),t_b(r)$.

The proposal to use galaxy number counts \cite{Romano:2007zz} to distinguish
between LTB models without cosmological constant and $\Lambda$CDM has been
recently studied both analytically \cite{Romano:2009qx,Romano:2009ej} and
numerically \cite{Celerier:2009sv} , showing how LTB models with a weak
central singularity could in principle not be distinguished even using both
the redshift spherical shell energy $mn(z)$ and $D_L(z)$.

In this paper we will take a different approach to the study of inhomogeneities \cite{Romano:2011mx,Romano:2012gk,Romano:2012kj},
and instead of proposing them as an alternative to dark energy, we will
consider their effects in presence of a cosmological constant, studying LTB
solutions which are only locally inhomogeneous whose geometry is very closed
to a $\Lambda$CDM model, showing how even small amplitude
inhomogeneities compatible
 with the amplitude of the curvature perturbation after inflation can lead to
important effects.
We in fact consider large scale inhomogeneities which today
are not linear and could be seeded by primordial curvature perturbations
compatible with CMB bounds.

Since the amplitude of the inhomogeneities we consider is very small,
corresponding to a few percent in terms of the density contrast,
the luminosity distance $D_L(z)$ is
not significantly affected as it can be seen in fig.(\ref{dzerr}), but the apparent cosmological observables derived
from the  $D_L(z)$ under the assumption of homogeneity are significantly
affected because they are sensitive to its derivatives.

We consider different types of models of local inhomogeneities and obtain
that local underdensity gives rise to apparent phantom behavior, while
local overdensity leads to apparent quintessence behavior.

Our study shows how observations of a quasi-$\Lambda$CDM universe with
compensated large scale inhomogeneities compatible with inflation predictions
 for curvature perturbations, if interpreted under the ``'wrong 'assumption
of homogeneity, can lead to the wrong conclusion of the presence of ``fake''
evolving dark energy, while only the cosmological constant is present
in reality.

\section{Deriving the exact LTB solution with a cosmological constant\label{ltb}}

The LTB solution can be written
as \cite{Lemaitre:1933qe,Tolman:1934za,Bondi:1947av} as
\begin{eqnarray}
\label{LTBmetric} %
ds^2 = -dt^2  + \frac{\left(R,_{r}\right)^2 dr^2}{1 + 2\,E(r)}+R^2
d\Omega^2 \, ,
\end{eqnarray}
where $R$ is a function of the time coordinate $t$ and the radial
coordinate $r$, $E(r)$ is an arbitrary function of $r$, and
$R_{,r}=\partial_rR(t,r)$.

The Einstein equations with dust and a cosmological constant give
\begin{eqnarray}
\label{eq2} \left({\frac{\dot{R}}{R}}\right)^2&=&\frac{2
E(r)}{R^2}+\frac{2M(r)}{R^3}+\frac{\Lambda}{3} \, , \\
\label{eq3} \rho(t,r)&=&\frac{2M,_{r}}{R^2 R,_{r}} \, ,
\end{eqnarray}
with $M(r)$ being an arbitrary function of $r$, $\dot
R=\partial_tR(t,r)$ and $c=8\pi G=1$ is assumed throughout the
paper. Since Eq. (\ref{eq2}) contains partial derivatives respect
to time only, its general solution can be obtained from the FLRW
equivalent solution by making every constant in the latter one an
arbitrary function of $r$.
Previous studies of LTB solutions in presence of a cosmological constant include \cite{omer,zecca,jef}.
The general analytical solution for a FLRW model with dust and cosmological
constant was obtained by Edwards \cite{Dilwyn} in terms of elliptic functions.
By an appropriate choice of variables and coordinates, we may extend it
to the LTB case thanks to the spherical symmetry of both LTB and FLRW models,
and to the fact that dust follows geodesics without being affected by
adjacent regions.
The Friedman equation for the scale factor $a_F(t)$ of a
pressureless FLRW universe with a cosmological constant has the
form: 
\be \left(\frac{\dot{a_F}}{a_F}\right)^2
=-\frac{k}{a_F^2}+\frac{\rho_0}{3a_F^3}+\frac{\Lambda}{3}\,. \ee
It is convenient to introduce the conformal time $\eta$ such that
$d\eta=dt/a_F$, 
in terms of which the solution satisfying the
initial Big-Bang condition $a_F(0)=0$ can then be
expressed as 
\bea
a_F(\eta)&=&\frac{\rho_0 }{3\phi(\frac{\eta}{2};{g_2,g_3})+k }\,; \quad
g_2=\frac{4}{3}k^2 \,,\quad
g_3=\frac{4}{27}\left(2 k^3 -\Lambda\rho_0^2\right)\,,
\eea where $\phi(x;g_2,g_3)$ is the Weierstrass elliptic function
satisfying the differential equation, 
\be
\left(\frac{d\phi}{dx}\right)^2=4\phi^3-g_2\phi-g_3\,.
\label{eqweir} \ee 
We note that in \cite{Dilwyn}, the curvature
parameter $k$ is normalized to $k=\pm1$, hence $k^2=1$. However,
for our purpose below we present the solution without normalizing
$k^2$ to unity.
We can now use this solution to construct a general solution of
the partial differential equation~(\ref{eq2}). First, we introduce
a new coordinate $\eta=\eta(t,r)$ and a variable $a$ by \bea
\left(\frac{\partial\eta}{\partial
t}\right)_r=\frac{r}{R}\equiv\frac{1}{a}\,, \label{etadef} \eea
and introduce new functions by \bea
\rho_0(r)\equiv\frac{6 M(r)}{r^3}\,,\quad
k(r)\equiv-\frac{2E(r)}{r^2}\,. \eea Then Eq. (\ref{eq2}) becomes
\be \left(\frac{\partial a}{\partial\eta}\right)^2 =-k(r)
a^2+\frac{\rho_0(r)}{3} a+\frac{\Lambda}{3} a^4\,, \ee where $a$ is now
regarded as a function of $\eta$ and $r$; $a=a(\eta,r)$. 
It should
be noted that the coordinate $\eta$, which is a generalization of
the conformal time in a homogeneous FLRW universe, has been only
implicitly defined by Eq.~(\ref{etadef}). The actual relation
between $t$ and $\eta$ can be obtained by integration $t=\int
a\,d\eta$ once $a(\eta,r)$ is known.
Inspired by the construction of the solution for the FLRW case, we finally get
\be
a(\eta,r)
=\frac{\rho_0(r)}{3\phi\left(\frac{\eta}{2 };g_2(r),g_3(r)\right)+k(r)}\,.
\ee
 
\section {Geodesic equations}
We adopt the same method developed in \cite{Romano:2009xw} to
solve the null geodesic equation written in terms of the
coordinates $(\eta,r)$. Instead of integrating differential
equations numerically, we perform a local expansion of the
solution around $z=0$ corresponding to the point $(t_0,0)$, or
equivalently $(\eta_0,0)$, where $t_0=t(\eta_0,0)$. The change of
variables from $(t,r)$ to $(\eta,r)$ permits us to have r.h.s. of
all equations in a fully analytical form, in contrast to previous
considerations of this problem which require a numerical
calculation of $R(t,r)$ from the Einstein equation~(\ref{eq2}).
Thus, this formulation is particularly suitable for derivation of
analytical results.

The luminosity distance for a central observer in the LTB
space-time as a function of the redshift $z$ is expressed as \be
D_L(z)=(1+z)^2 R\left(t(z),r(z)\right) =(1+z)^2
r(z)a\left(\eta(z),r(z)\right) \,, \ee where
$\Bigl(t(z),r(z)\Bigr)$ or $\Bigl((\eta(z),r(z)\Bigr)$ is the
solution of the radial geodesic equation as a function of $z$. The
past-directed radial null geodesics is given by \bea \label{geo1}
\frac{dt}{dr}
=-\frac{R_{,r}(t,r)}{\sqrt{1+2E(r)}} \,. \eea
In terms of $z$, Eq. (\ref{geo1}) takes the form
 \cite{Celerier:1999hp}:
\begin{eqnarray}
{dr\over dz}&=&{\sqrt{1+2E(r(z))}\over {(1+z){\dot R}_{,r}[r(z),t(z)]}} \,,
\nonumber\\
{dt\over dz}&=&-{R_{,r}[r(z),t(z)]\over {(1+z){\dot
R}_{,r}[r(z),t(z)]}} \,. \label{eq:35}
\end{eqnarray}
The inconvenience of using the $(t,r)$ coordinates is that there
is no exact analytical solution for $R(t,r)$. So the r.h.s. of
Eqs. (\ref{eq:35}) cannot be evaluated analytically, but we are
required to find a numerical solution for $R$ first
\cite{Hellaby:2009vz}, and then to integrate numerically the
differential equations, which is quite an inconvenient and
cumbersome procedure. Alternatively one may derive a local
expansion of $R(t,r)$ around $(t_0,0)$, corresponding to the
central observer, and use it Eqs.~(\ref{eq:35}). But one would
need to expand it to a higher order in $z$ in order to maintain
the accuracy at high redshifts.

For this reason, it is useful for many numerical and analytical
applications to write the geodesic equations in terms of the
coordinates $(\eta,r)$. It follows from the definition
(\ref{etadef}) that \bea t(\eta,r)
&=&t_b(r)+\int^{\eta}_{0}a(\eta^{'},r) d\eta^{'} \, , \label{tsol}
\eea hence, \bea dt&=&a(\eta,r)d\eta+\left(\int^{\eta}_{0}
\frac{\partial a(\eta^{'},r)}{\partial r}
d\eta^{'}+t_b^{'}(r)\right) dr \,. \eea Partial derivatives are
transformed using the relations: \be \label{partial}
\left(\frac{\partial}{\partial
t}\right)_r=a^{-1}\left(\frac{\partial}{\partial
\eta}\right)_r,~~~\left(\frac{\partial}{\partial r}\right)_t=
\left(\frac{\partial}{\partial
r}\right)_{\eta}-a^{-1}\left(\frac{\partial t}{\partial
r}\right)_{\eta}\left(\frac{\partial}{\partial \eta}\right)_r \,.
\ee 
Then Eqs. (\ref{eq:35}) take the form: 
\bea \label{geo3}
\frac{d \eta}{dz} &=&-\frac{\partial_r
t(\eta,r)+F(\eta,r)}{(1+z)\partial_{\eta}F(\eta,r)}
\equiv p(\eta,r) \,,\\
\label{geo4} \frac{dr}{dz}
&=&\frac{a(\eta,r)}{(1+z)\partial_{\eta}F(\eta,r)} \equiv
q(\eta,r) \,, \eea where \be F(\eta,r)\equiv \frac{\
R_{,r}}{\sqrt{1+2E(r)}}=
\frac{1}{\sqrt{1-k(r)r^2}}\left[\partial_r (a(\eta,r) r)
-a^{-1}\partial_{\eta} (a(\eta,r) r)\, \partial_r t(\eta,r)\right]
\,. \ee
It is important that the functions $p,q,F$ have explicit
analytical forms.


\section{Apparent cosmological observables and ``fake'' dark energy}
In this section we will briefly introduce the concept of apparent observables,
which are deduced from observations assuming a flat $\Lambda$CDM model.
We start from observing that in a flat FLRW model \cite{Sahni:1999gb}
 there are simple relations between the Hubble parameter $H(z)$, the luminosity distance  $D_L(z)$, the cosmic deceleration $q(z)$ and the effective equation of state of dark energy $w_{DE}(z)$ :
\bea
H^{FLRW}(z)=\left[\frac{d}{dz}\left(\frac{D^{FLRW}_L(z)}{1+z}\right)\right]^{-1}\,.
\\
Q^{FLRW}(z)=\frac{d}{dz}\left(\frac{D^{FLRW}_L(z)}{1+z}\right)=(H^{FLRW}(z))^{-1}\,,
\\
q^{FLRW}(z)=-1-\frac{d {\ln}(Q^{FLRW}(z))}{d{\ln}(1+z)}=q^{FLRW}(D_L(z))\,,
\\
w^{FLRW}_{DE}(z) =\frac{(2(1+z)/3)\,d{\rm ln}H^{FLRW}\,/dz - 1}{1 - (H_0/H^{FLRW})^2 \,,
\Omega_{0m}(1+z)^3}\,\,.
\eea
We will use the above relations to define apparent observables in terms of the observed luminosity distance according to
\bea
H^{app}(z)=\left[\frac{d}{dz}\left(\frac{D^{obs}_L(z)}{1+z}\right)\right]^{-1}\,,
\\
Q^{app}(z)=\frac{d}{dz}\left(\frac{D^{obs}_L(z)}{1+z}\right)=(H^{app}(z))^{-1}\,,
\\
q^{app}(z)=-1-\frac{d {\ln}(Q^{app}(z))}{d{\ln}(1+z)}=q^{app}(D^{obs}_L(z))\,,
\\
w^{app}_{DE}(z) =\frac{(2(1+z)/3)\,d{\rm ln}H^{app}\,/dz - 1}{1 - (H_0/H^{app})^2 
\Omega_{0m}(1+z)^3}\,\,.
\eea
where $q^{app}(z)$ is the apparent cosmic deceleration parameter, $w^{app}_{DE}(z)$ the apparent equation of state of dark energy and $Q^{app}(z)$ is an auxiliary function introduced of mathematical convenience.
Apparent observables are deduced from the observed luminosity distance assuming the same functional relations which apply to the case of a homogeneous and isotropic universe described by a FLRW metric.
We will apply these definitions of apparent observables assuming the observed luminosity distance corresponds to the case of a central observer in a $\Lambda$LTB space. 
If the Universe is really inhomogeneous the apparent observables
above will include the errors due to ignoring the inhomogeneity,
which could for example be mistaken as dark energy with
a redshift dependent equation of state, that is, we may be fooled
by `fake' dark energy.

In this paper in addition to the above, we will also consider
$Om(z)$~\cite{Sahni:2008xx,ZC08}, a diagnostic which can be used to
distinguish $\Lambda$CDM from other DE models without directly
involving the cosmic equation of state, \bea Om^{app}(x) \equiv
\frac{h^2(x)-1}{x^3-1},~~ x=1+z~,~h(x) = H^{app}(x)/H_0~.
\label{eq:om} \eea

\section{Effects of local inhomogeneities on apparent observables}

In order to make a connection between the LTB model and a universe
with primordial curvature perturbations from inflation, we
introduce the following metric which describes a spherically
symmetric space-time after inflation on scales much exceeding the
Hubble scale: 
\be ds^2=-dt^2 + a_F^2(t)e^{2\zeta(r)}(dr^2+r^2d\Omega^2)\,. 
\label{zeta}
\ee 
According to the inflationary
scenario, $\zeta(r)$ is just a local, space-dependent number of
$e$-folds $N$ produced during inflation (up to a constant which may be
absorbed into $a_F$). This relation which constitutes the basis of
the so-called $\delta N$ formalism was first obtained in \cite{S82} 
in case of a single field inflation, and then
generalized to multiple field inflation in \cite{S85,SS96};
see also \cite{Lyth:2004gb} for further consideration.

This metric in general can describe a stage dominated by any form of the matter
with any equation of state. But it is valid only for inhomogeneities whose
characteristic scale $L$ is much greater than the Hubble horizon scale,
$HL\gg1$. In other words, it is the metric valid at leading order 
in the spatial gradient expansion~\cite{Lyth:2004gb}.
Since we have $L=a_F\ell$ where $\ell$
is the comoving scale, while $H$ scales as $a_F^{-2}$ during the 
radiation-dominated stage or $a_F^{-3/2}$ during the matter-dominated stage, 
$HL=Ha_F\ell=\dot a_F\ell$ is a decreasing function of time.
So for a given comoving scale of inhomogeneities $\ell$, the metric (\ref{zeta}) 
is valid only at an early stage of the universe when $\dot a_F\ell\gg1$.

On the other hand, the LTB metric is valid on any scales but only after
the universe has become matter-dominated.
Fortunately, for sufficiently large scale inhomogeneities,
say those whose scale is 1/10 of the current Hubble radius,
which is the case for specific examples studied below,
there is a sufficiently wide overlap of time during which both
metrics are valid, after the recombination of hydrogens at the
redshift $z\sim10^3$ until the redshift $\sim100$. 

The important point is that this identification allows us to make a direct
connection between the primordial curvature perturbations encoded in $\zeta(r)$
and the function $k(r)$ appearing in the LTB solution. In this way we can justify
 the size of the present day local large scale inhomogeneity we consider by
 relating it to early time curvature perturbations, making our model more realistic.
Since the metric (\ref{zeta}) is valid on super-Hubble scales,
 at the time of the matching the perturbations seeding our present day
 inhomogeneity were super-horizon, but today has become sub-horizon.

Matching the metric (\ref{zeta}) with the LTB metric
at the stage when both are valid as descibed above,
we obtain the following relations:
\bea
R&=&a_F(t)e^{\zeta(r)}r\,, 
\\
1+2 E(r)&=&[1+r \zeta'(r) ]^2 \,. \label{zetaErel}
\eea
We note that the first equality is only approximate.
It ignores corrections of the order $(HL)^{-2}$.
But the second equality is exact in the sense that
the amplitude of $\zeta$ does not have to be small, though
it is small in reality. At the linear approximation, this reduces to
\be 
k(r)=-2\frac{\zeta'(r)}{r}\,. 
\ee 
In particular, $k(0)=-2\zeta''(0)$. 
Note that the LTB metric (\ref{LTBmetric}) or the
relation (\ref{zetaErel}) is invariant under the change of the
radial coordinate $r\to \bar r=g(r)$, where $g(r)$ is an arbitrary
function of $r$ as long as it is monotonic in $r$. 
We will choose the coordinates in which $\rho_0=constant$.
This approach has the goal to establish some quantitative bounds on 
the size of a spherically symmetric inhomogeneity which may surround us 
and is compatible with the inflationary predictions. Using the above relations 
we can in fact directly relate the function $k(r)$, which in the coordinates  
we chose completely determine an LTB model, with the primordial curvature 
perturbations produced by inflation. This makes our models more realistic 
since we will study inhomogeneities which are compatible with inflation 
predictions.

Motivated by observations we consider the curvature perturbation
$\zeta(r)$ of the amplitude $\sim5\times 10^{-5}$.
Specifically we study the four different types of inhomogeneities,
\bea
\mbox{Type I}^{-}\,:\ k(r)=\frac{A}{r_0^2}
[l(r)-l'(0) r e^{-r/r_0}]; && H_0r_0=0.1\,,\ A = 10^{-4}\,,
\ \Delta= 0.02\,, \\
\mbox{Type I}^{+}\,:\ k(r)=\frac{A}{r_0^2}
[l(r)-l'(0) r e^{-r/r_0}]; && H_0r_0=0.1\,,\ A=-10^{-4}\,,
\ \Delta= 0.02\,, \\
\mbox{Type II}^{-}:\ \zeta(r)=A[l(r)-l'(0) r e^{-r/r_0}];
&& H_0r_0=0.2\,,\ A=5\times 10^{-5}\,,\ \Delta=0.05\,, \\ \label{type2m}
\mbox{Type II}^{+}:\ \zeta(r)=A[l(r)-l'(0) r e^{-r/r_0}];
&& H_0r_0=0.2\,,\ A=-5\times 10^{-5}\,,\ \Delta=0.05\,, \label{type2p}
\eea
where the function $l(r)$ is defined as
\be
l(r)=\left[\tanh{\left(\frac{H_0(r-r_0)}{\Delta}\right)}-1\right]\,.
\ee
In all cases, the cosmological constant is assumed to be the same
as the one implied by the best fit $\Lambda$CDM model
corresponding to $\Omega_{\Lambda}=0.7$,
and $H_0=H^{app}(z=0)$ is adjusted to the observed Hubble constant.

We have chosen these four different inhomogeneity profiles because
they correspond to compensated inhomogeneities, in the sense that
they asymptotically approach a flat, homogeneous $\Lambda$CDM model
and correspond, respectively, to
\begin{itemize}
\item Type I$^{+}$ : central overdense region.
\item Type I$^{-}$ : central underdense region.
\item Type II$^{+}$ : intermediate overdense region.
\item Type II$^{-}$ : intermediate underdense region.
\end{itemize}
An important feature of type I models is that they are by
construction regular at the center, since the linear term in
the series expansion at the center is removed, avoiding the
cusp singularity which would otherwise arise.
Type II models
are also regular at the center, i.e. $k(r)$ and $k'(r)$ do not diverge, though it is not immediately apparent. It can be seen
by expanding $\zeta(r)$ in qqs.(\ref{type2m},\ref{type2p}) around $r=0$.

As seen from Figs.~(\ref{krI}-\ref{omzI}) for
type I$^{\pm}$ models,
the presence of a local underdensity gives rise to apparent phantom behavior,
while that of a local overdense region to apparent quintessence behavior.
For a shell-like underdensity or overdensity region,
Figs.~(\ref{krII}-\ref{omzII}) for type II$^\pm$ also indicate
that an underdensity shell mimics phantom behavior,
while an overdensity shell mimics quintessence behavior.

Since all the models considered here describe compensated inhomogeneities,
they show both phantom and quiescence behaviors, but we can see
that the sign of the variation of $w_{DE}^{app}$ with respect
to $w_{true}=-1$ is roughly the same as the sign of the density contrast.

Our results give a semi-realistic example of inhomogeneities
compatible with observations which, if interpreted in the framework
of a flat and inhomogeneous spacetime, can lead to the wrong conclusion
that there exists dark energy with an evolving equation of state,
while in reality there is only a cosmological constant.

The relation between density profiles and the
free parameters of the LTB metric is quite difficult
and elaborate, as shown for example in \cite{Mustapha:2000bf,Sussman:2010ew}.
One useful way to analyze the relation between the profile
of $k(r)$ with that of $\rho(r)$ is to consider the linear theory limit.
On sufficiently large scales where the amplitude of $\zeta$ is small,
$\zeta\ll1$, and its spatial variation is sufficiently smooth, $r|\zeta'(r)|\ll1$,
$\zeta$ is equal to the conserved comoving curvature perturbation in linear theory,
which it is essentially equal to the minus of the Newton potential $\Psi$
in the Newton gauge. At the matter-dominated stage, $\Psi=-(3/5)\zeta$. 
So we can estimate the resulting density profile
for a given profile of $\zeta$ from the Poisson equation, 
\bea
\delta\rho\propto \Delta\Psi\propto \frac{3}{10} r^{-2}\frac{d(r^3k(r))}{dr}
\eea
where we have used the fact that in the limit $r|\zeta'(r)|\ll1$ eq.(28) holds.
Although this is
valid only at linear order in the strict sense, it gives the correct qualitative
behavior of the density profile. This indeed shows that the sign of $k(r)$ is
an important factor but not always the dominant factor in the determination
of the density profile.

\section{Analytical derivation of $q_0^{app}$ and $w_0^{app}$ }

The calculation of the central value of the apparent cosmological
 observables requires to find an analytical expression for the
 right hand side of the geodesics equations.
For this purpose we expand the relevant functions as
\bea
t(\eta,r)&=&A_0(\eta)+A_1(\eta)r+\frac{1}{2}A_2(\eta)r^2 +\cdots\,,\\
\eta(z)&=&\eta_0+\eta_1 z+\eta_2 z^2 +\cdots\,,\\
r(z)&=&r_1 z+r_2 z^2 +\cdots\,,
\eea
to get 
\bea
q_0^{app}&=&-\frac{2 
\left(r_1 a_{,r}+\eta_1 a_{,\eta}\right)}{a}-\frac{2 r_2}{r_1}-3\,,
\\
w_0^{app}&=&-\frac{4 r_1 
\left(r_1 a_{,r}+\eta_1 a_{,\eta}\right)+(7 r_1+4 r_2) a}{3 r_1 a(1-H_0^2 \Omega_M(r_1 a)^2)}
\,.
\eea
The solution of the geodesics equations leads to
\bea
r_1&=&\frac{a^0}{a^0_{,\eta}}\,,
\\
r_2&=&\frac{a \left(a_{,\eta} A_1'(\eta_0)+2 A_1(\eta_0) a_{,\eta\eta}-2
   a_{,\eta}^2+a_{,r} a_{,\eta}\right)-2 A_1(\eta_0) a_{,\eta}^2
+\left(a_{,\eta\eta}-2 a_{,\eta r}\right) a^2}{2 a_{,\eta}^3}
\,,\\
\eta_1&=&-\frac{a^0+A_1(\eta_0)}{a^0_{,\eta}}\,.
\eea

We can then also expand the energy density around the center,
\bea
\rho(\eta,r)=\frac{\rho_0}{a(\eta ,0)^3}
+\frac{r \rho_0\left[A_1(\eta ) a_{,\eta}(\eta ,0)
-4 a(\eta ,0) a_{,r}(\eta ,0)\right]}{a(\eta ,0)^5}+O\left(r^2\right)\,.
\eea
In order to avoid a central singularity the term linear in $r$ should vanish, 
and from this condition we obtain
\bea
A_1(\eta)&=&\frac{4 a(\eta ,0) a_{,r}(\eta ,0)}{a_{,\eta}(\eta ,0)}\,.
\eea
Using this equation we finally get
\bea
q_0^{app}&=&\frac{a_{,r} \left(9 a_{,\eta}^2-4 a a_{,\eta\eta}\right)+a_{,\eta}
   \left(a_{,\eta}^2-a \left(2 a_{,\eta r}+a_{,\eta\eta}\right)\right)}{a_{\eta}^3}
\,,\\
w_0^{app}&=&\frac{a_{,r} \left(8 a a_{,\eta\eta}-18 a_{,\eta}^2\right)+a_{,\eta}
   \left(2 a \left(2 a_{,\eta r}+a_{,\eta\eta}\right)-a_{,\eta}^2\right)}{3
   a_{,\eta} \left(H_0^2 \Omega_M a^4 -a_{,\eta}^2\right)}\,,
\eea
where the right-hand sides are evaluated at $(\eta,r)=(\eta_0,0)$.
It may be worth mentioning that
the formulas derived so far are general in the sense 
they do not depend on the explicit form of the solution.

We can also define 
\bea
q_0 &= -&\frac{\ddot{a}(t_0,0)\dot{a}(t_0,0)}{\dot{a}(t_0,0)^2}\,,
\eea
where the derivative respect to $t$ is denoted with a dot, and is 
calculated using the analytical solution $a(\eta,r)$ and the 
derivative respect to $\eta$,
\be
\dot{a}=\partial_{t}a=\frac{\partial_{\eta}a}{a}\,.
\ee
It is interesting to observe that because of the regularity 
condition we have imposed at the center, we have
\bea
q_0^{app}&=&q_0\,,
\eea
which can be verified using the analytical solution both 
for the case of vanishing and non vanishing cosmological constant.

Using again the condition $k_1=0$ we can now substitute the analytical solution to get the final results expressed directly in terms of observables:
\bea
q^{app}_0 
&=& \frac{3}{2}\Omega_M-1 + 2\, \tilde{\zeta''}(0) ,
\\ \label{q0app}
w^{app}_0 &=& -1+ \frac{4}{3(1-\Omega_M)}\tilde{\zeta''}(0)\,, \label{w0app}
\eea 
where we have used 
\bea
a_0 &=& \frac{L^2 \rho_0}{\zeta''(0) L^2+3 \phi_0 }\,, \\
H_0 &=& -\frac{3 \phi'_0}{2 L^3 \rho_0}\,,\\
\phi_0&=&\phi \left(\frac{\eta_0}{2 L};\frac{16}{3} \zeta''(0)^2L^4,-\frac{4}{27} \left(16 \zeta''(0)^3+ \Lambda\rho _0^2\right)L^6\right)\,,\\
\phi_0'&=&\partial_{x}\phi \left(x;\frac{16}{3} \zeta''(0)^2L^4,-\frac{4}{27} \left(16 \zeta''(0)^3+ \Lambda\rho _0^2\right)L^6\right)
\Biggr|_{x=\frac{\eta_0}{2 L}}\,,\\
\Lambda &=& 3 (1-\Omega_M) H_0^2 \,,\\
\tilde{\zeta''}(0)&=&\frac{1}{(a_0 H_0)^2}\zeta''(0)\,,\\
\rho_0&=&3 a_0^3 \Omega_M H_0^2\,.
\eea
As expected the above formulae reduce to the $\Lambda CDM$ 
case in the central flat limit,
\bea
k_0&=-2 \zeta''(0)&=0\,, \\
q_0^{app}&=q_0^{\Lambda CDM} &= \frac{3}{2}\Omega_M-1\,, \\
w_0^{app}&=w_0^{\Lambda CDM} &= -1\,.
\eea
As a confirmation that large scale inhomogeneities look like fake
dark energy we can also verify that the relation between 
$q_0^{app}$ and $w_0^{app}$ is the same as in the case of 
an FLRW model with dark energy:
\bea
q_0^{FLRW}&=&\frac{3}{2}\Omega_m-1+\frac{3}{2}(1+w_0^{DE})(1-\Omega_M)\,, \\
q_0^{app}&=&\frac{3}{2}\Omega_m-1+\frac{3}{2}(1+w_0^{app})(1-\Omega_M)\,.
\eea 
It should be noted that the above relations are general since they do not depend on the particular type of inhomogeneity profile, follow directly from the definition of $w^{app}$ and $q^{app}$ and have been derived to show that a large scale inhomogeneity self-consistently mimick evolving dark energy.

\begin{table}
	\centering
		\begin{tabular} {|l|l|l|}
			\hline
			Type & $q_0^{app}$ & $w_0^{app}$ \\
			\hline
			$\Lambda CDM$ & -0.58 & -1  \\
			\hline
			I  & -0.56001 & 0.981482\\
			\hline
			II  & -0.599999& -1.01852\\
			\hline
			III  & -0.579866& -0.999876\\
			\hline
			IV  & -0.580134 & -1.00012 \\
			\hline
		\end{tabular}
		\caption{The apparent value of $q_0$ and $w_0$ is given using the analytical formula in eq.( \ref{q0app},\ref{w0app}). In all the five models $\Omega_{\Lambda}=0.72$. The $\Lambda CDM$ case is reported for reference for reference to the standard cosmological model.}
\end{table}

\section{Conclusions}

We have investigated how the presence of a local inhomogeneity could affect
the apparent equation of state of dark energy under the ``wrong'' assumption
of a homogeneous FLRW background, which is commonly used in
interpreting astrophysical observations in $\Lambda$CDM models.
Our calculation shows how phantom and quintessence behaviors can be produced
for compensated underdense or overdense regions.
The presence of a local underdensity gives rise to apparent phantom
behavior, while that of a local overdense region to apparent
quintessence behavior.

Our results give a semi-realistic example of inhomogeneities with
the amplitude compatible with inflationary predictions which,
if interpreted in the framework of a flat and
inhomogeneous spacetime, can lead to the wrong conclusion of the presence
of dark energy with an evolving equation of state.
In general, a local inhomogeneity can lead to a confusion between local
 gravitational redshift and cosmological redshift due to the expansion of
the Universe.

Recent analysis of observational data \cite{Shafieloo:2009ti} could support
the existence of a local underdense region, but which may not be of
compensated type as the one we considered here.
We will investigate in a future work what could be the constraints on the
size and density contrast of such a void based on observational data.

\acknowledgments
AER is supported by MEXT Grant-in-Aid for the global COE program
 at Kyoto University,
"The Next Generation of Physics, Spun from Universality and Emergence", by the UDEA research project IN615CE and the dedicacion exclusiva program.
MS is supported in part by JSPS Grant-in-Aid for Scientific
Research (A) No.~21244033, and by JSPS
Grant-in-Aid for Creative Scientific Research No.~19GS0219.
AS acknowledges RESCEU hospitality as a visiting professor. He was also
partially supported by the grant RFBF 11-02-00643 and by the Scientific
Programme ``Astronomy'' of the Russian Academy of Sciences.

\appendix
\section{Calculating the density contrast}

In the text, we have carried out all our calculations in the coordinates
$(\eta,r)$
since this allows to take full advantage of the existence of an analytical
solution. But if we are interested in the radial profile of
a quantity on a fixed time-slice $t=$constant, we need to go
back to the coordinates $(t,r)$. Below we carry this out for
the density contrast, $\delta=(\rho(t,r)-\rho(t,\infty))/\rho(t,\infty)$,
where our LTB model is assumed to approach a flat FLRW universe as $r\to\infty$.

We need to introduce the inverse of the function defined in eq.~(\ref{tsol}),
i.e., we need to express $\eta$ as a function of $(t,r)$, $\eta=v(t,r)$,
from
\bea
t&=&u(\eta,r)=t_b(r)+\int^{\eta}_{0}a(\eta^{'},r) d\eta^{'} \,,
\eea
such that
\be
u(v(t,r),r)=t\,.
\ee
The value of $v(t,r)$ can be evaluated numerically by solving for $x$
 the equation
\be
u(x,r)=t\,.
\ee
The function $\eta=v(t_0,r)$ thus obtained
is plotted for the different models in Figs.~\ref{etaI} and \ref{etaII}.
As it can be seen, $\eta=v(t_0,r)$ varies substantially in
the region of inhomogeneity, while it levels off to a constant
far from the inhomogeneity.

The energy density in the coordinates $(t,r)$ is given by
\be
\rho(t,r)=\frac{2\, \partial_rM}{R^2(t,r)\partial_r R(t,r)}\,.
\ee
But since the analytical solution is given in terms of $\eta$
 we have another expression,
\be
\rho(\eta,r)=\frac{2\, \partial_r M}{ a(\eta,r)^2 r^2 [\partial_r (a(\eta,r)r)
 - a^{-1}\partial_{\eta} (a(\eta,r) r) \partial_r t] }\,.
\ee
Then the density contrast on the hypersurface $t=t_0$ is given by
\be
\delta(t_0,r)=\frac{\rho(\eta(t_0,r),r)
-\rho(\eta(t_0,\infty),\infty)}{\rho(\eta(t_0,\infty)),\infty)}\,,
\ee
where
\bea
t_0=t(\eta_0,0)\,.
\eea
The density contrast is plotted in Fig.~\ref{deltaI}
for type I$^\pm$ inhomogeneities
and Fig.~\ref{deltaII} for type II$^\pm$ inhomogeneities.

\newpage

\begin{center}
\begin{figure}[t]
\includegraphics[height=55mm,width=80mm]{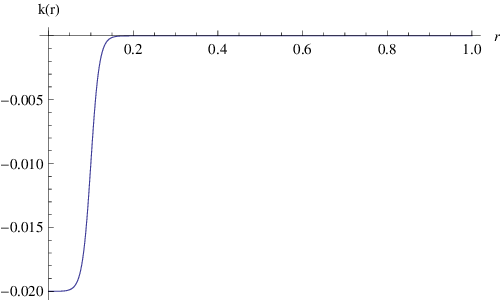}
\includegraphics[height=55mm,width=80mm]{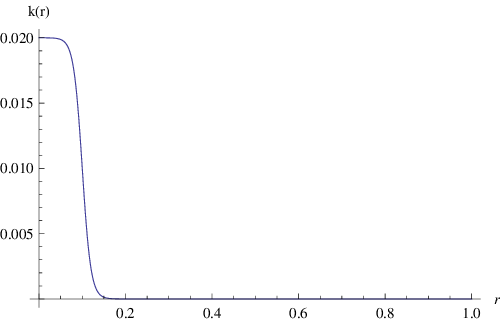}
\caption{$k(r)$ in units of $H_0^2$ is plotted
for inhomogeneity of type I$^{-}$ and I$^{+}$. $r$ is in units of $H_0^{-1}$.}
\label{krI}
\end{figure}

\begin{figure}[h]
\includegraphics[height=55mm,width=80mm]{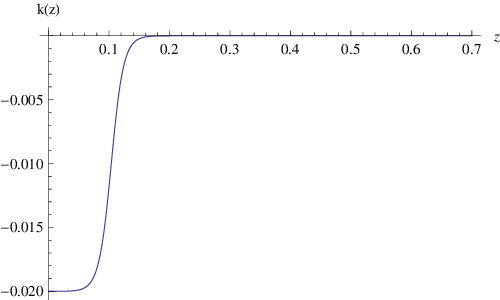}
\includegraphics[height=55mm,width=80mm]{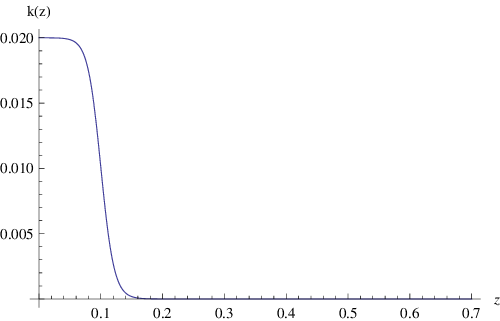}
\caption{$k(z)$ is plotted for inhomogeneity of type I$^{-}$ and I$^{+}$.
}
\label{kzI}
\end{figure}

\begin{figure}[h!]
\includegraphics[height=55mm,width=80mm]{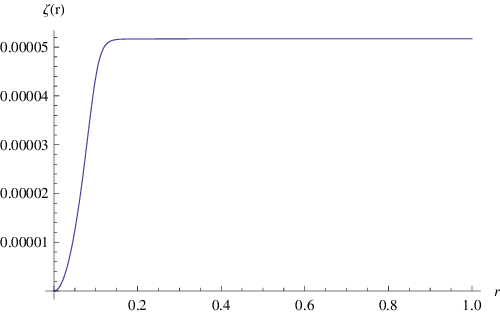}
\includegraphics[height=55mm,width=80mm]{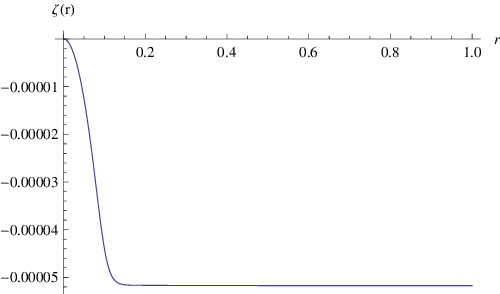}
\caption{$\zeta(r)$ is plotted for inhomogeneity of type I$^{-}$ and I$^{+}$.}
\label{zetaI}
\end{figure}
\end{center}

\begin{center}
\begin{figure}[t]
\includegraphics[height=55mm,width=80mm]{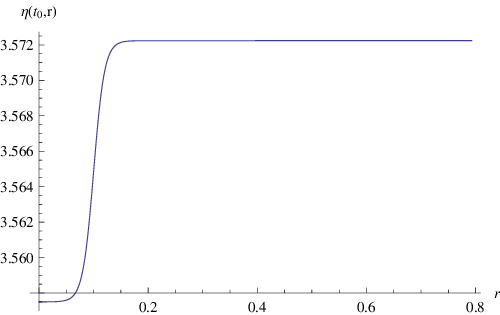}
\includegraphics[height=55mm,width=80mm]{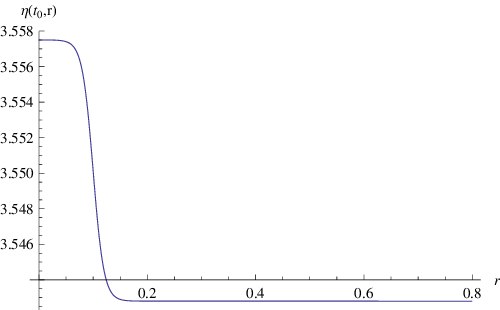}
\caption{ $\eta=v(t_0,r)$ in units of $H_0^{-1}$ is plotted
for inhomogeneity of types I$^{-}$ and I$^{+}$.}
\label{etaI}
\end{figure}

\begin{figure}[h]
\includegraphics[height=55mm,width=80mm]{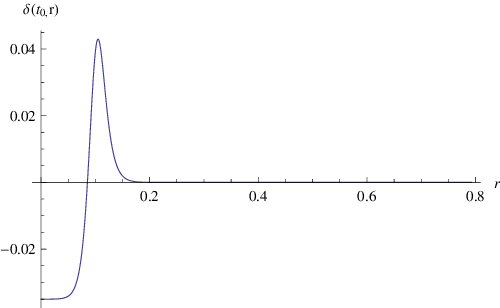}
\includegraphics[height=55mm,width=80mm]{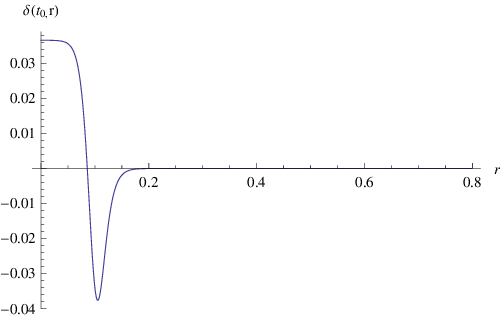}
\caption{ $\delta(t_0,r)$ is plotted as a function of $r$
for inhomogeneity of types I$^{-}$ and I$^{+}$.}
\label{deltaI}
\end{figure}

\begin{figure}[h!]
\includegraphics[height=55mm,width=80mm]{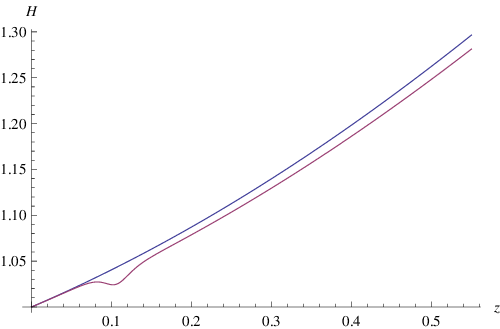}
\includegraphics[height=55mm,width=80mm]{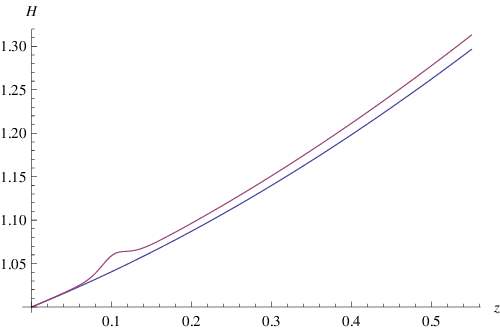}
\caption{ $H^{app}(z)$ is plotted for inhomogeneity of types I$^{-}$ and I$^{+}$.}
\label{HzI}
\end{figure}
\end{center}


\newpage

\begin{center}
\begin{figure}[t]
\includegraphics[height=55mm,width=80mm]{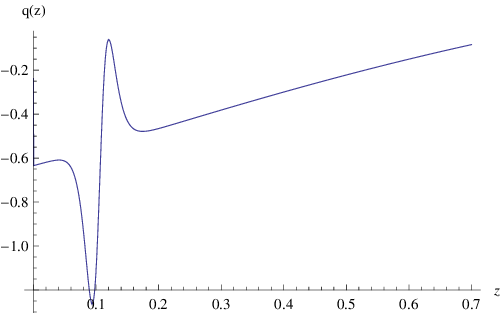}
\includegraphics[height=55mm,width=80mm]{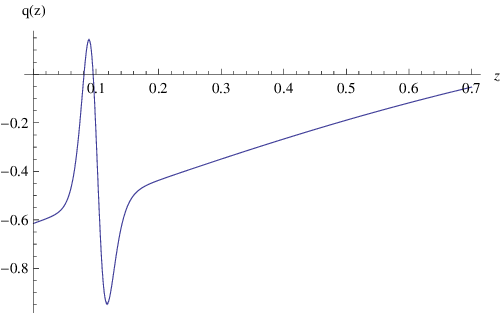}
\caption{ $q^{app}(z)$ is plotted for inhomogeneity of types I$^{-}$ and I$^{+}$.}
\label{qzI}
\end{figure}

\begin{figure}[h]
\includegraphics[height=55mm,width=80mm]{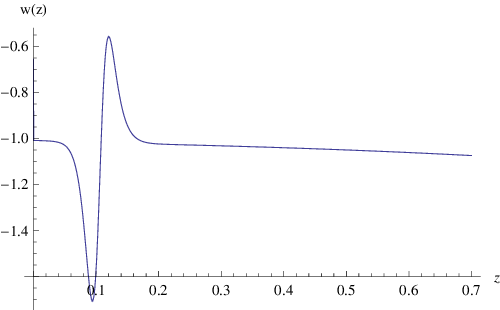}
\includegraphics[height=55mm,width=80mm]{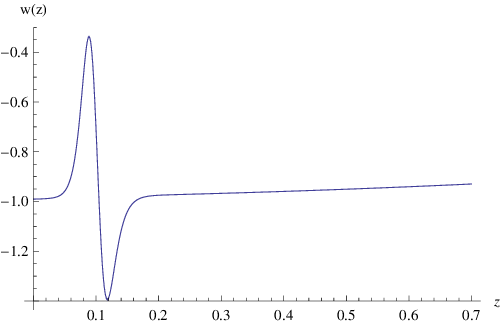}
\caption{ $w^{app}_{DE}(z)$ is plotted for inhomogeneity of types I$^{-}$ and I$^{+}$.}
\label{wzI}
\end{figure}

\begin{figure}[h!]
\includegraphics[height=55mm,width=80mm]{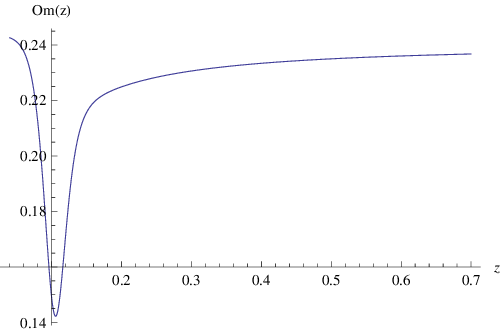}
\includegraphics[height=55mm,width=80mm]{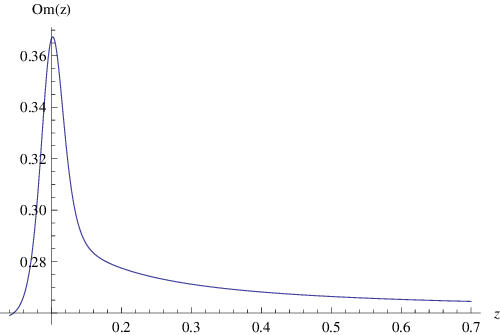}
\caption{ $Om^{app}(z)$ is plotted for inhomogeneity of types I$^{-}$ and I$^{+}$.}
\label{omzI}
\end{figure}
\end{center}

\newpage
\begin{center}
\begin{figure}[t]
\includegraphics[height=55mm,width=80mm]{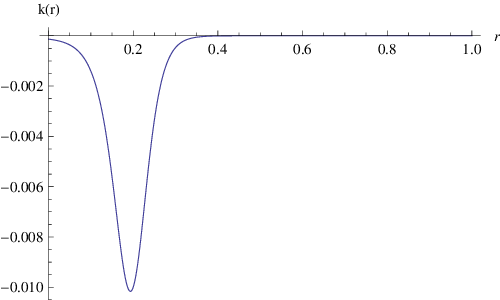}
\includegraphics[height=55mm,width=80mm]{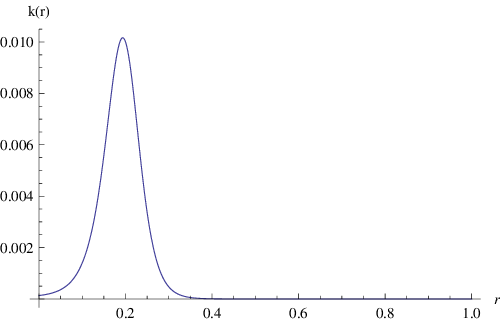}
\caption{ The same as Fig.~\ref{krI} but
for inhomogeneity of types II$^{-}$ and II$^{+}$.}
\label{krII}
\end{figure}

\begin{figure}[h]
\includegraphics[height=55mm,width=80mm]{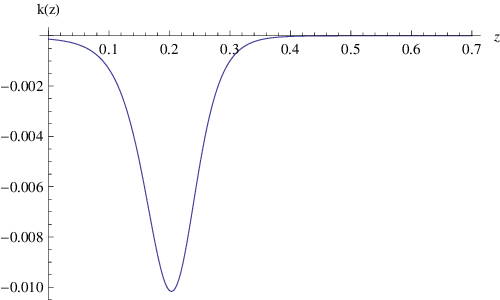}
\includegraphics[height=55mm,width=80mm]{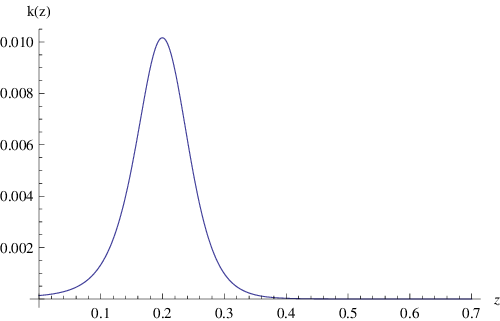}
\caption{ The same as Fig.~\ref{kzI} but
 for inhomogeneity of types II$^{-}$ and II$^{+}$.}
\label{kzII}
\end{figure}

\begin{figure}[h!]
\includegraphics[height=55mm,width=80mm]{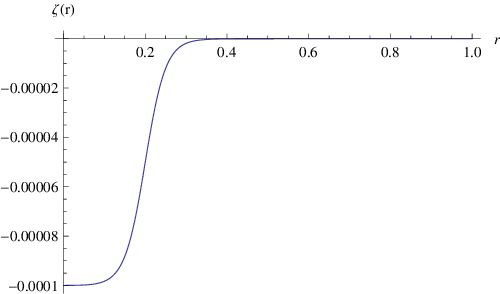}
\includegraphics[height=55mm,width=80mm]{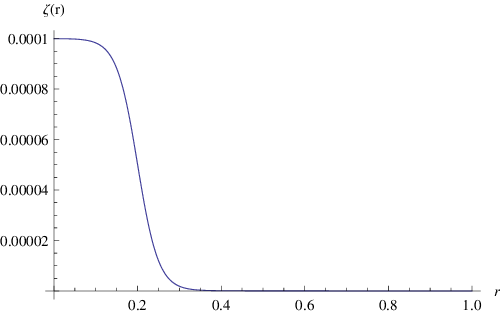}
\caption{ The same as Fig.~\ref{zetaI} but
for inhomogeneity of types II$^{-}$ and II$^{+}$.}
\label{zetaII}
\end{figure}
\end{center}

\begin{center}
\begin{figure}[t]
\includegraphics[height=55mm,width=80mm]{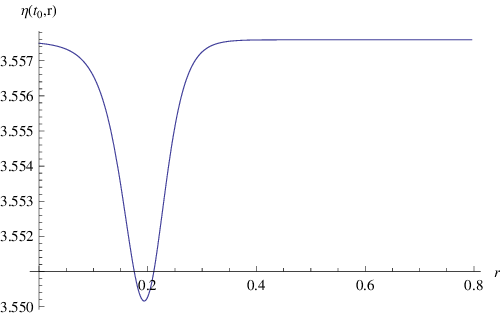}
\includegraphics[height=55mm,width=80mm]{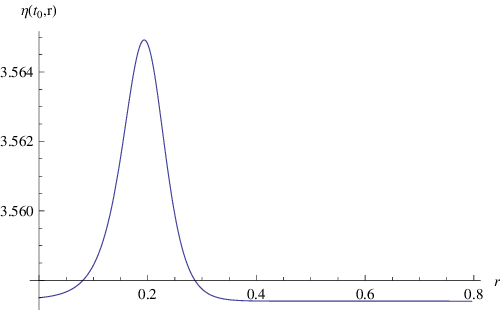}
\caption{The same as Fig.~\ref{etaI} but
for inhomogeneity of types II$^{-}$ and II$^{+}$.}
\label{etaII}
\end{figure}

\begin{figure}[h]
\includegraphics[height=55mm,width=80mm]{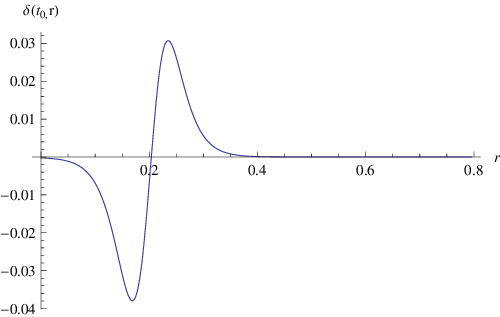}
\includegraphics[height=55mm,width=80mm]{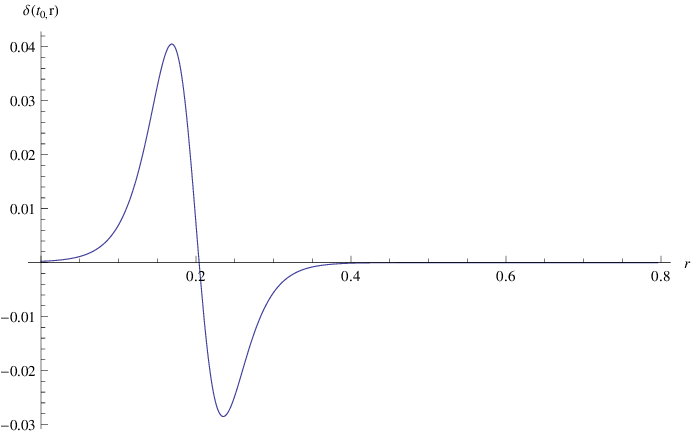}
\caption{The same as Fig.~\ref{deltaI} but
for inhomogeneity of types II$^{-}$ and II$^{+}$.}
\label{deltaII}
\end{figure}

\begin{figure}[h!]
\includegraphics[height=55mm,width=80mm]{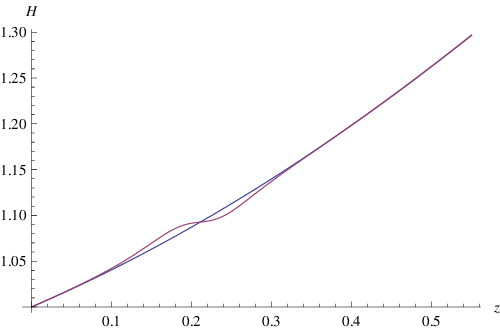}
\includegraphics[height=55mm,width=80mm]{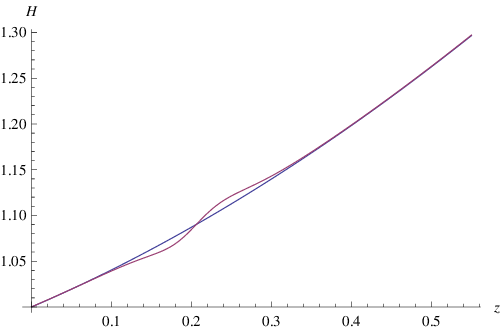}
\caption{The same as Fig.~\ref{HzI} but
 for inhomogeneity of types II$^{-}$ and II$^{+}$.}
\label{HzII}
\end{figure}
\end{center}

\begin{center}
\begin{figure}[t]
\includegraphics[height=55mm,width=80mm]{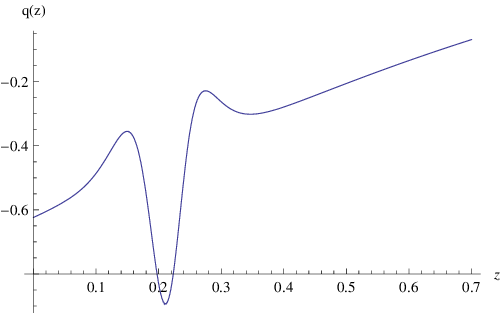}
\includegraphics[height=55mm,width=80mm]{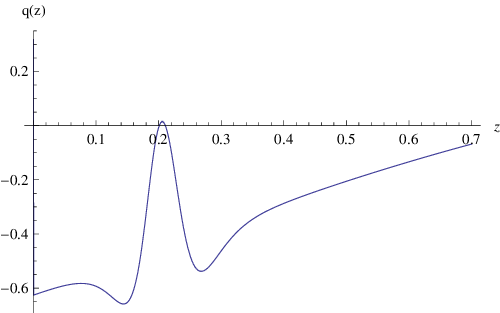}
\caption{ The same as Fig.~\ref{qzI} but
 for inhomogeneity of types II$^{-}$ and II$^{+}$.}
\label{qzII}
\end{figure}

\begin{figure}[h]
\includegraphics[height=55mm,width=80mm]{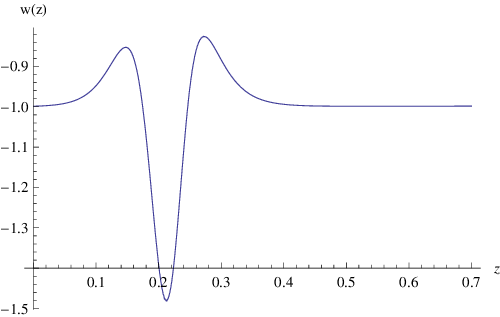}
\includegraphics[height=55mm,width=80mm]{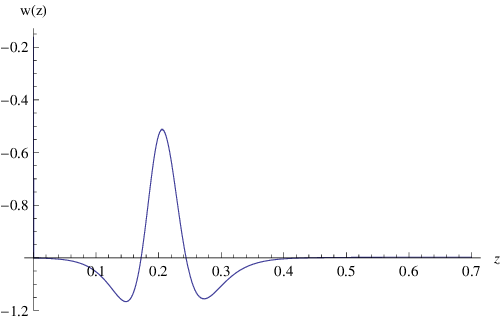}
\caption{The same as Fig.~\ref{wzI} but
for inhomogeneity of types II$^{-}$ and II$^{+}$.}
\label{wzII}
\end{figure}

\begin{figure}[h!]
\includegraphics[height=55mm,width=80mm]{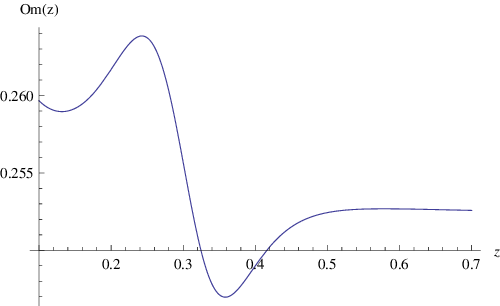}
\includegraphics[height=55mm,width=80mm]{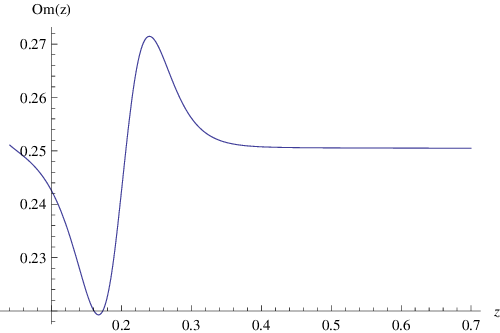}
\caption{The same as Fig.~\ref{omzI} but
 for inhomogeneity of types II$^{-}$ and II$^{+}$.}
\label{omzII}
\end{figure}

\begin{figure}[h!]
\includegraphics[height=55mm,width=80mm]{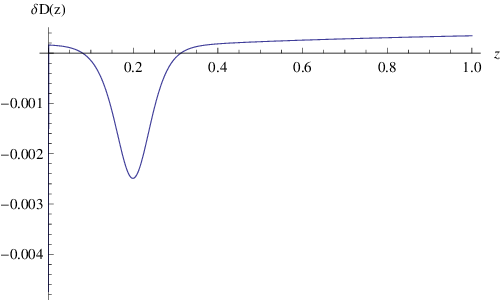}
\caption{$\delta D_L(z)=[D_L^{\Lambda CDM}(z)-D_L(z)]/D^{\Lambda CDM}_L(z)$, the relative difference between the luminosity distances,  is plotted for model II$^{+}$. As it can be seen the large scale inhomogeneities considered have a very small effect.}
\label{dzerr}
\end{figure}
\end{center}


\begin{thebibliography}{99}


\bibitem{Riess:1998cb}
  A.~G.~Riess {\it et al.}  [Supernova Search Team Collaboration],
  Astron.\ J.\  {\bf 116}, 1009 (1998)
  [arXiv:astro-ph/9805201].


\bibitem{Tonry:2003zg}
  J.~L.~Tonry {\it et al.}  [Supernova Search Team Collaboration],
  Astrophys.\ J.\  {\bf 594}, 1 (2003)
  [arXiv:astro-ph/0305008].

\bibitem{Knop:2003iy}
  R.~A.~Knop {\it et al.}  [The Supernova Cosmology Project Collaboration],
  Astrophys.\ J.\  {\bf 598}, 102 (2003)
  [arXiv:astro-ph/0309368].

\bibitem{Barris:2003dq}
  B.~J.~Barris {\it et al.},
  Astrophys.\ J.\  {\bf 602}, 571 (2004)
  [arXiv:astro-ph/0310843].

\bibitem{Riess:2004nr}
  A.~G.~Riess {\it et al.}  [Supernova Search Team Collaboration],
  Astrophys.\ J.\  {\bf 607}, 665 (2004)
  [arXiv:astro-ph/0402512].

\bibitem{Perlmutter:1999np}
S.~Perlmutter {\it et al.}  [Supernova Cosmology Project
Collaboration],
``Measurements of Omega and Lambda from 42 High-Redshift Supernovae,''
Astrophys.\ J.\  {\bf 517}, 565 (1999) [arXiv:astro-ph/9812133].

\bibitem{WMAP2003}
  C.~L.~Bennett {\it et al.},
  Astrophys.\ J.\ Suppl.\  {\bf 148}, 1 (2003)
  [arXiv:astro-ph/0302207];

\bibitem{Spergel:2006hy} 
  D.~N.~Spergel {\it et al.}  [WMAP Collaboration],
  Astrophys.\ J.\ Suppl.\  {\bf 170}, 377 (2007)
  [astro-ph/0603449].


\bibitem{Nambu:2005zn}
  Y.~Nambu and M.~Tanimoto,
  arXiv:gr-qc/0507057.

\bibitem{Kai:2006ws}
  T.~Kai, H.~Kozaki, K.~i.~nakao, Y.~Nambu and C.~M.~Yoo,
  Prog.\ Theor.\ Phys.\  {\bf 117}, 229 (2007)
  [arXiv:gr-qc/0605120].

\bibitem{Romano:2006yc}
  A.~E.~Romano,
  Phys.\ Rev.\  D {\bf 75}, 043509 (2007)
  [arXiv:astro-ph/0612002].

\bibitem{Chung:2006xh}
  D.~J.~H.~Chung and A.~E.~Romano,
  Phys.\ Rev.\ D {\bf 74}, 103507 (2006)
  [arXiv:astro-ph/0608403].


\bibitem{Alexander:2007xx} 
  S.~Alexander, T.~Biswas, A.~Notari and D.~Vaid,
  JCAP {\bf 0909}, 025 (2009)
  [arXiv:0712.0370 [astro-ph]].

\bibitem{Alnes:2005rw}
  H.~Alnes, M.~Amarzguioui and O.~Gron,
  Phys.\ Rev.\ D {\bf 73}, 083519 (2006)
  [arXiv:astro-ph/0512006].

\bibitem{Romano:2007zz}
  A.~E.~Romano,
  Phys.\ Rev.\  D {\bf 76}, 103525 (2007)
  [arXiv:astro-ph/0702229].

\bibitem{Romano:2009qx} 
  A.~E.~Romano,
  JCAP {\bf 1001}, 004 (2010)
  [arXiv:0911.2927 [astro-ph.CO]].

\bibitem{Romano:2009ej} 
  A.~E.~Romano,
  JCAP {\bf 1005}, 020 (2010)
  [arXiv:0912.2866 [astro-ph.CO]].
	

\bibitem{Celerier:2009sv} 
  M.~-N.~Celerier, K.~Bolejko and A.~Krasinski,
  Astron.\ Astrophys.\  {\bf 518}, A21 (2010)
  [arXiv:0906.0905 [astro-ph.CO]].


\bibitem{Romano:2011mx} 
  A.~E.~Romano and P.~Chen,
  JCAP {\bf 1110}, 016 (2011)
  [arXiv:1104.0730 [astro-ph.CO]].
	
	
\bibitem{Romano:2012gk} 
  A.~E.~Romano and P.~Chen,
  arXiv:1207.5572 [astro-ph.CO].
	
\bibitem{Romano:2012kj} 
  A.~E.~Romano and P.~chen,
  Int.\ J.\ Mod.\ Phys.\ D {\bf 20}, 2823 (2011)
  [arXiv:1208.3911 [gr-qc]].


\bibitem{Lemaitre:1933qe}
  G.~Lemaitre,
  Annales Soc.\ Sci.\ Brux.\ Ser.\ I Sci.\ Math.\ Astron.\ Phys.\ A {\bf 53}, 51 (1933).
\bibitem{Tolman:1934za}
  R.~C.~Tolman,
  Proc.\ Nat.\ Acad.\ Sci.\  {\bf 20}, 169 (1934).
\bibitem{Bondi:1947av}
  H.~Bondi,
  Mon.\ Not.\ Roy.\ Astron.\ Soc.\  {\bf 107}, 410 (1947).
\bibitem{omer}
G C Omer, Proc Natl Acad Sci 53, 1 (1965)


\bibitem{zecca}
A Zecca, Nuovo Cimento B 106, 413 (1991)

\bibitem{jef}
J E Felten and R Isaacman, Rev Mod Phys 58, 689 (1986).


\bibitem{Dilwyn}
D. Edwards, Monthly Notices of the Royal Astronomical Society, 159, 51 (1972).


\bibitem{Romano:2009xw} 
  A.~E.~Romano and M.~Sasaki,
  Gen.\ Rel.\ Grav.\  {\bf 44}, 353 (2012)
  [arXiv:0905.3342 [astro-ph.CO]].

\bibitem{Celerier:1999hp}
  M.~N.~Celerier,
  Astron.\ Astrophys.\  {\bf 353}, 63 (2000)
  [arXiv:astro-ph/9907206].

\bibitem{Hellaby:2009vz}
  C.~Hellaby,
  PoS {\bf ISFTG}, 005 (2009)
  [arXiv:0910.0350 [gr-qc]].


\bibitem{Sahni:1999gb} 
  V.~Sahni and A.~A.~Starobinsky,
  Int.\ J.\ Mod.\ Phys.\ D {\bf 9}, 373 (2000)
  [astro-ph/9904398].


\bibitem{Sahni:2008xx} 
  V.~Sahni, A.~Shafieloo and A.~A.~Starobinsky,
  Phys.\ Rev.\ D {\bf 78}, 103502 (2008)
  [arXiv:0807.3548 [astro-ph]].



\bibitem{ZC08}
  C.~Zunckel and C.~Clarkson, Phys.\ Rev.\ Lett. {\bf 101}, 181301
  (2008) [arXiv:0807.4304 [astro-ph]].

\bibitem{S82}
  A.~A.~Starobinsky, Phys.\ Lett.\ B  {\bf 117}, 175 (1982).

\bibitem{S85}
  A.~A.~Starobinsky, JETP Lett. {\bf 42}, 152 (1985).

\bibitem{SS96}
  M.~Sasaki and E.~D.~Stewart, Progr.\ Theor.\ Phys. {\bf 95}, 71
  (1996) [arXiv:astro-ph/9507001]].

\bibitem{Mustapha:2000bf} 
  N.~Mustapha and C.~Hellaby,
  Gen.\ Rel.\ Grav.\  {\bf 33}, 455 (2001)
  [astro-ph/0006083].

\bibitem{Lyth:2004gb}
  D.~H.~Lyth, K.~A.~Malik and M.~Sasaki,
  JCAP {\bf 0505}, 004 (2005)
  [astro-ph/0411220].


\bibitem{Sussman:2010ew} 
  R.~A.~Sussman,
  Class.\ Quant.\ Grav.\  {\bf 27}, 175001 (2010)
  [arXiv:1005.0717 [gr-qc]].

\bibitem{Shafieloo:2009ti}
  A.~Shafieloo, V.~Sahni and A.~A.~Starobinsky,
  Phys.\ Rev.\  D {\bf 80}, 101301 (R) (2009)
  [arXiv:0903.5141 [astro-ph.CO]].










%
%
%
%




%
%
%

%
%
%






%
%




%

%

%


\end{thebibliography}
\end{document}